\documentclass[journal=jpclcd,manuscript=letter]{achemso}

\usepackage[version=3]{mhchem} 
\usepackage[T1]{fontenc}       

\usepackage{amssymb}
\usepackage{graphicx}
\usepackage{color}
\usepackage{float}

\author{Abraham Nitzan}
\affiliation{Department of Chemistry, University of Pennsylvania, Philadelphia, Pennsylvania 19104, USA}
\affiliation{School of Chemistry, Tel Aviv University, Tel Aviv, 69978, Israel}
\author{Michael Galperin}
\affiliation{Department of Chemistry and Biochemistry, University of California at San Diego, La Jolla, CA 92093, USA}
\email{migalperin@ucsd.edu}
\phone{+1-858-246-0511}

\title{Kinetic Schemes in Open Interacting Systems}

\begin{document}

\begin{tocentry}
{\centering\includegraphics[width=5cm]{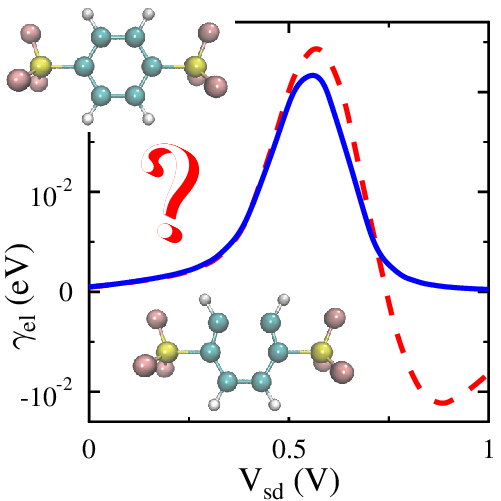}}\\*[0.2cm]
\begin{flushleft}
\large No phonon runaway:\\ kinetic scheme vs. NEGF.
\end{flushleft}
\end{tocentry}

\begin{abstract}
We discuss utilization of kinetic schemes for description of open interacting systems, 
focusing on vibrational energy relaxation for an oscillator coupled to a nonequilibirum electronic bath.
Standard kinetic equations with constant rate coefficients are obtained 
under the assumption of timescale separation between system and bath,
with the bath dynamics much faster than that of the system of interest.
This assumption may break down in certain limits and we show that ignoring this
may lead to qualitatively wrong predictions.
Connection with more general, nonequilibrium Green's function (NEGF)
analysis, is demonstrated. Our considerations are illustrated
within generic molecular junction models with electron-vibration coupling.
\end{abstract}

{\em Introduction.}
Development of experimental techniques on the nanoscale has made studies of
single molecule junctions possible. These experiments yield unique possibilities to
explore physical and chemical properties of molecules by measuring their responses
to external perturbations. Following experimental advances, there was rapid development
of theoretical approaches. Today a variety of techniques, ranging from diagrammatic expansions
(such as, e.g., nonequilibrium Green function (NEGF)~\cite{HaugJauho_2008,StefanucciVanLeeuwen_2013}, 
quantum master equation~\cite{LeijnseWegewijsPRB08,GrifoniEPJB13}, 
and Hubbard NEGF~\cite{ChenOchoaMGJCP17,MiwaChenMGSciRep17}) 
to approximate treatments of strongly correlated systems
(e.g., dynamical mean field theory~\cite{WernerRMP14} and beyond~\cite{KatsnelsonLichtensteinRubtsovRMP18})
and to numerically exact methods (e.g., renormalization group techniques~\cite{PaasckeRoschWolflePRB04,Anders2008,UedaJPhysSocJpn08,MedenNJP14}
and continuous time quantum Monte-Carlo~\cite{CohenPRL14,CohenGullReichmanMillisPRL15,GullCohenPRB18}) are available.
Implementation of such schemes, particularly the numerically exact approaches, 
is often expensive and their applications for simulations of realistic systems is limited.

At the same time, simple kinetic schemes have been widely
and successfully utilized in description of rate phenomena in open molecular systems
(for example, donor-bridge-acceptor (DBA) molecular complexes)~\cite{MiglioreNitzanACSNano11,EinaxDierlNitzanJPCC11,MiglioreNitzanJACS13,EinaxNitzanJPCC14,CravenNitzanPNAS16,CravenNitzanJCP17,ChenCravenNitzanJCP17,CravenNitzanPRL17}.
Such schemes lead to description of system states connected by rate processes 
whose Markov limit description provides
``rate coefficients'' which enter into the kinetic description (master equation) of the system evolution.
Such Markov limit descriptions rely on timescale separation between the observed 
system evolution and dynamic processes that determine the rates, the latter usually involves the 
dynamics of relaxation in the bath. 
Obviously the details of such kinetic schemes depend on the way system-bath separation is defined 
and used. The general practice dictated by balance between simplicity and rigor, takes ``the system'' 
to be comprised by the observed variables together (when possible) with other variables 
whose inclusion makes the dynamics Markovian. This practice should be exercised with caution 
because even if conditions for timescale separations are established in a given range, 
they may become invalid in other domains of operation. 
Such situations are well known in classical dynamics. For example, transition state theory (TST) 
of molecular rate processes assumes that molecular degrees of freedom 
except the reaction coordinate are at thermal equilibrium (and therefore can be taken 
to be part of the thermal environment). TST breaks down when the observed rate is of the order of, 
or faster than, the rate of thermal relaxation in the molecule, as demonstrated in 
the Kramers theory of activated rate processes~\cite{Kramers1940}. 

Importantly, even if the assumed timescale separation holds near equilibrium it might fail far from it.
The reason is that systems interacting with their equilibrium surrounding remain within 
the energetic domain of thermal energy, while systems coupled to non-equilibrium environments,
e.g., under optical illumination or voltage bias, may be driven to energy domains where
timescale separation does not hold. Thus, while situations of the first kind (such as activated 
barrier crossing)
are well understood and documented, mathematically equivalent 
circumstances have been often overlooked. One such case is the vibrational dynamics of molecules 
adsorbed at or bridging between metallic interfaces due to coupling to the thermal electronic baths. 
Molecular vibrational motion is sensitive to the electronic occupation of the molecule,
which in turn is affected by the molecule and metal electronic structure, their mutual coupling
and the junction voltage bias.
A common approximation, equivalent to the fast bath assumption discussed above, 
is to disregard the effect of vibrational dynamics on the electronic subsystem, 
representing the latter by a thermal electronic bath or, for a biased (current carrying) junction, 
by the corresponding steady-state electronic distribution, 
assumed unaffected by the vibrational process. This level of description, 
which effectively takes the molecular electronic degrees of freedom as part of 
the (generally non-equilibrium) electronic bath~\cite{SegalPRB15,EspositoPRB18,SubotnikPRB18},
has been recently used to discuss bias induced vibrational 
instabilities~\cite{BrandbygePRL11,SegalPCCP12,ThossPeskinNL18}.
While the limitations of such treatments are sometimes pointed out~\cite{ThossPeskinNL18},
in other publications they are ignored.
Indeed, such instabilities were recently claimed\cite{FortiVazquezJPCL18} to be 
generic properties of wires whose conduction is dominated by distinct electronic resonances 
(or, in the language of  Ref.~\cite{FortiVazquezJPCL18} by ``separated'' electronic states).

It should be noted that in general, zero order treatments of the kind described above
are known to violate conservation laws~\cite{BaymKadanoffPR61,BaymPR62}.
Thus, notwithstanding their usefulness in many applications, such treatment should be regarded 
with caution, in particular when unusual behavior are observed. 
For example, the observation of negative sign of vibrational dissipation rate at an apparent 
steady-state of a molecular junction should not be regarded an indicator of a true vibrational instability 
in the system, but (like in linear stability analysis of non-linear differential equations) as an indication 
of failure of the underlying assumptions that lead to such result. It should be emphasized that 
(again, as in linear stability analysis) such analysis can be useful as an indicator that a real 
stable state exists elsewhere (which in a real anharmonic molecule may or may not lie beyond 
a bond breaking threshold). Still, many low order treatments~\cite{BrandbygePRL11,SegalPCCP12}
of vibrational instabilities in harmonic bridge models of molecular junctions leave the reader 
with the message that the observed ``runaway behavior'' describes the full physical behavior.

\begin{figure}[htbp]
\centering\includegraphics[width=\linewidth]{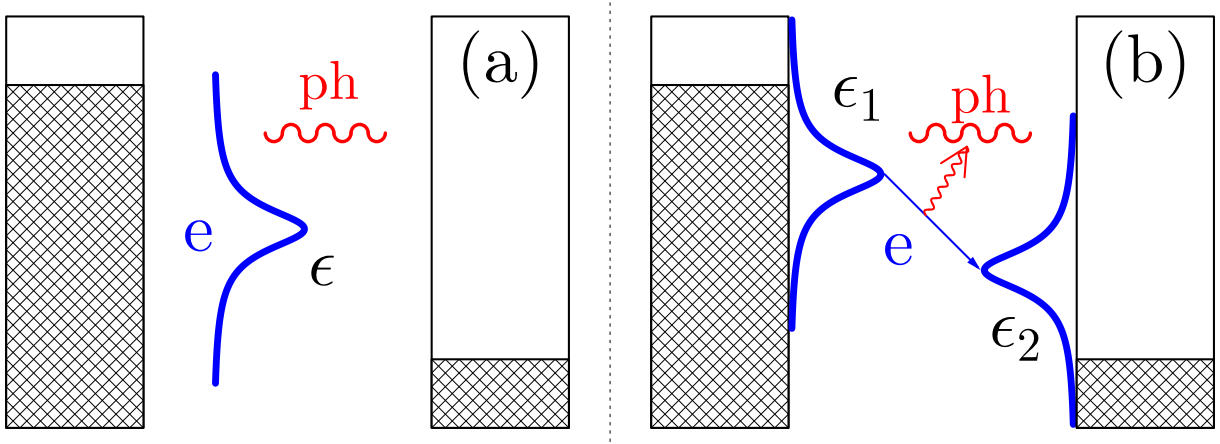}
\caption{\label{fig1}
Molecular junction with electron-phonon interaction. Shown are models for
(a) single level junction with polaronic coupling and
(b) two-level junction with non-adiabatic coupling.
Note that setup (b) favors phonon heating, which is maximized when the level spacing 
$\epsilon_1-\epsilon_2$ is close to the vibrational frequency.
} 
\end{figure}

Exact numerical solutions~\cite{schinabeck_hierarchical_2018} are obviously capable of exploring 
the correct physical picture. Here we show that an approximate self-consistent treatment 
that does not violate conservation laws can already avoid the qualitative pitfalls of a linear theory. 
We consider simple junction models with electron-vibration coupling (see Fig.~\ref{fig1}),
treated within the nonequilibrium Green's function (NEGF) theory. 
In this framework the consequence of interaction between a system of interest 
(here the vibrational mode) and its environment (here the electronic subsystem) 
enters through self-energy terms that (a)~directly describe relaxation and driving of the system 
of interest by its environment and (b)~modify the Green functions that enter into the definition of 
these self-energies.  A full calculation must therefore be self-consistent and take into account 
the mutual influence between system and environment, namely the effect of environment 
on the system as well as the back-action from the system on the environment. 
We show that the basis of the zero order approximation, the assumption of timescale separation
between molecular electronic and vibrational degrees of freedom
(i.e. ability to neglect back action of molecular vibration on electronic bath),
does not hold automatically, 
and in fact has a limited range of validity.  Consequently, while regimes of significantly enhanced 
vibrational heating can be found in biased junctions with electron-phonon coupling 
(and heating transient spikes may occur following sudden parameter change), 
instabilities identified as appearance of negative vibrational dissipation rate do not occur.
In our consideration below, the
molecular vibration (system) is weakly coupled to the electronic degrees of freedom (bath),
which is the usual setup in considerations of system and bath separation.
We stress that even in this favorable situation kinetic considerations may lead to qualitative failures.

Below, after introducing the model, we discuss its
general treatment using NEGF and its connection to simple kinetic considerations.
We consider the steady-state of such model under voltage bias and illustrate failures of standard kinetic description 
within numerical examples.
 
{\em Model and method.}
We consider a junction consisting of a molecular bridge $M$ coupled to two contacts, $L$ and $R$
(Fig~\ref{fig1}).
Besides electronic degrees of freedom, a molecular vibration, modeled as a harmonic oscillator
of frequency $\omega_0$ coupled to the molecular electronic subsystem, is included. 
The contacts are reservoirs of free charge carriers, each at its own equilibrium.
The model Hamiltonian is
\begin{align}
 \hat H &= \hat H_M + \sum_{K=L,R}\bigg(\hat H_K + \hat V_{KM}\bigg)
 \\
 \label{HM}
 \hat H_M &= \sum_{m_1m_2\in M} H^M_{m_1m_2}\hat d_{m_1}^\dagger\hat d_{m_2} 
 + \omega_0\hat a^\dagger\hat a 
 \\ &
 + \sum_{m_1,m_2\in M} U_{m_1m_2}\bigg(\hat a+\hat a^\dagger\bigg)
    \hat d_{m_1}^\dagger\hat d_{m_2}
\nonumber \\
 \hat H_K &= \sum_{k\in K}\epsilon_k\hat c_k^\dagger\hat c_k
 \\
 \hat V_{KM} &= \sum_{k\in K}\sum_{m\in M}\bigg( V_{km}\hat c_k^\dagger\hat d_m + H.c. \bigg)
\end{align}
where $\hat d_m^\dagger$ ($\hat d_m$) and $\hat c_k^\dagger$ ($\hat c_k$) creates (annihilates)
electron in level $m$ of the bridge and state $k$ of contacts, respectively.
$\hat a^\dagger$ ($\hat a$) creates (annihilates) vibrational quanta.
$V_{km} = V_{k1}$ when $k\in L$ and $V_{k2}$ when $k\in R$ is molecule-contact transfer
matrix element and $U_{m_1m_2}$ is electron-phonon coupling strength.
Below we consider two special cases of this Hamiltonian: 
A single bridge level with polaronic coupling to the vibrational mode, whereupon the last term in 
Eq.~(\ref{HM})  takes the form $U (\hat a+\hat a^\dagger)\hat d^\dagger \hat d$ (Fig.~\ref{fig1}a)
and a bridge comprising two coupled electronic levels, each coupled  to its respective lead, 
with electron-vibration coupling of the form 
$U (\hat a+\hat a^\dagger)(\hat d_2^\dagger \hat d_1+\hat d_1^\dagger \hat d_2)$.

We treat the electron-vibration coupling, last term in Eq.(\ref{HM}),  within standard 
diagrammatic technique. According to the rules for building conserving approximations~\cite{BaymKadanoffPR61,BaymPR62}
one starts from the Luttinger-Ward functional~\cite{LuttingerWardPR60,StefanucciVanLeeuwen_2013},
whose functional derivatives with respect to the electron and phonon (vibration) Green functions yield
the electron self-energy due to coupling to vibrations, $\Sigma^{(ph)}$, 
and the phonon self-energy due to coupling to electronic degrees of freedom, $\Pi^{(el)}$, respectively.  
For our discussion it is important to stress that the electron and phonon
Green functions in the functional are full (dressed) functions, with back-action of electrons on vibration
and vice versa taken into account. Explicit expressions at second order of the diagrammatic technique 
in electron-phonon interaction are~\cite{ParkMG_FCS_PRB11,GaoMGJCP16_1}
\begin{align}
\label{Sigma}
\Sigma^{(ph)}_{m_1m_2}(\tau_1,\tau_2) &= i\, D(\tau_1,\tau_2)\,
\mbox{Tr}\bigg[\mathbf{U}\,\mathbf{G}(\tau_1,\tau_2)\,\mathbf{U}\bigg]
\\
\label{Pi}
\Pi^{(el)}(\tau_1,\tau_2) &= -i\,\mbox{Tr}\bigg[\mathbf{U}\,\mathbf{G}(\tau_1,\tau_2)\,\mathbf{U}\,\mathbf{G}(\tau_2,\tau_1)\bigg]
\end{align}
where the $\mbox{Tr}[\ldots]$ is over electronic degrees of freedom in $M$ and
\begin{align}
 G_{m_1m_2}(\tau_1,\tau_2) &= 
  -i\langle T_c\, \hat d_{m_1}(\tau_1)\,\hat d_{m_2}^\dagger(\tau_2)\rangle
  \\
  D(\tau_1,\tau_2) &= -i\langle T_c\,\hat a(\tau_1)\,\hat a^\dagger(\tau_2)\rangle
\end{align}
are the electron and phonon (vibration) Green functions
(here $T_c$ is the Keldysh contour ordering operator and $\tau_{1,2}$ are the contour variables). 
Below (for simplicity and to compare with previous studies) we will consider the quasiparticle
limit for the phonon Green function~\cite{Mahan_1990}. 
Solving together the coupled Eqs.~(\ref{Sigma}) and (\ref{Pi}) constitutes the self-consistent
Born approximation (SCBA)\footnote{Note that within SCBA, the electron self-energy contains also the Hartree term, which comes from an additional contribution to the Luttinger-Ward functional and is responsible for 
shift of electronic levels due to the interaction. For relatively weak electron-vibration coupling
the shift is small and can be disregarded.}.
Dynamical characteristics are obtained by projecting these Keldysh functions onto real time.
Lesser and greater projections of the self-energies (\ref{Sigma})-(\ref{Pi}) describe respectively 
in- and out-fluxes into the corresponding degree of freedom due to its coupling to 
the other degrees of freedom in the system, while the retarded projection describes dissipation induced 
by the interaction. These projections are related by
\begin{equation}
\label{relation_Pi}
 \Pi^{(el)\, >}(t_1,t_2)-\Pi^{(el)\, <}(t_1,t_2) = \Pi^{(el)\, r}(t_1,t_2)-\Pi^{(el)\, a}(t_1,t_2)
\end{equation}  
where $\Pi^{(el)\, a}(t_1,t_2)=[\Pi^{(el)\, r}(t_2,t_1)]^{*}$ is the advanced projection and $t_{1,2}$ are 
physical times corresponding to the contour variables $\tau_{1,2}$. A similar relation holds for 
the electron self-energies obtained as projections of (\ref{Sigma}) onto the physical time.

At steady state, when correlation functions depend on time differences, one can Fourier transform 
(\ref{relation_Pi}). The right side of the expression is identified to be the vibrational dissipation rate
due to coupling to electronic degrees of freedom
\begin{equation}
\label{gamma_el}
 \gamma_{el}(\omega) = i\bigg( \Pi^{(el)\, >}(\omega)-\Pi^{(el)\, <}(\omega) \bigg)
\end{equation}
which at quasiparticle limit should be taken at $\omega=\omega_0$.
The energy flux between the electronic and vibrational subsystems can be expressed by either of the two fluxes~\cite{MGNitzanRatner_heat_PRB07,Datta_1995}
\begin{align}
\label{Iel_ph}
I^{(el)}_{ph} &= -\int_0^\infty \frac{d\omega}{2\pi}\bigg( 
 \Pi^{(el)\, <}(\omega)\, D^{>}(\omega) - \Pi^{(el)\, >}(\omega)\, D^{<}(\omega),
 \bigg)
\\
\label{Iph_el}
I^{(ph)}_{el} &= \int\frac{dE}{2\pi}\,
 \mbox{Tr}\bigg[\Sigma^{(ph)\, <}(E)\, \mathbf{G}^{>}(E) - \Sigma^{(ph)\, >}(E)\, \mathbf{G}^{<}(E)
 \bigg]
\end{align}
which can be shown, by substituting for the self-energies $\Sigma$ and $\Pi$
the corresponding projections of Eqs.~(\ref{Sigma}) and (\ref{Pi}), respectively,
to be equal in magnitude and opposite in sign. 
These fluxes are caused by the electron-phonon interaction.
Eq.~(\ref{Iel_ph}) expresses the energy flux
(in terms of vibrational quanta) into the vibrational system, while (\ref{Iph_el}) expresses the
flux for population redistribution between energy levels of the electronic subsystem.
Because of charge conserving character of electron-phonon interaction this flux vanishes, 
which at the quasi-particle limit leads to~\cite{FranssonMGPCCP11}
\begin{equation}
\label{N_w0}
 N_{\omega_0} =i\,\Pi^{(el)\, <}(\omega_0)/\gamma_{el}(\omega_0)
\end{equation}
Here $N_{\omega_0}=\langle\hat a^\dagger\hat a\rangle$ is the nonequilibrium average
phonon population. 

As discussed above, zero order treatments that lead to
standard kinetic schemes for this problem assume timescale separation between
electronic and vibrational equilibration times (usually treating the vibrational subsystem as
much slower than its electronic counterpart), thus disregarding back action of the phonon on 
the electronic subsystem.
Mathematically this is manifested by disregarding contribution to the electron self-energy due 
to coupling to vibration, Eq~(\ref{Sigma}), 
and employing the resulting zero order electronic Green functions in evaluation of phonon self-energy (\ref{Pi}).
While the argument of timescale separation seems reasonable, it may lead to erroneous
predictions.  
We note in passing that within diagrammatic perturbation theory, 
substituting full (dressed) Green function with the bare one in
the Luttinger-Ward functional leads to violation of conservation laws in the 
system~\cite{KadanoffBaym_1962}.
In the Markov limit, when self-energies simplify into transition rates,
such substitution corresponds to statement that rates are kept constant irrespective
to actual state of the system.

Below we illustrate some consequences of breakdown of such time scale separation assumption 
with numerical examples for the model junctions shown in Fig.~\ref{fig1}.
Model (a) comprises a single molecular electronic level coupled to the two metal
electrodes and to a single vibration, with $H^M_{mm}\to\epsilon$ and $U_{mm}\to U$.
Model (b) involves two molecular levels and one vibrational mode with
$H^M_{m_1m_2}=\delta_{m_1,m_2}\epsilon_{m_1} - (1-\delta_{m_1,m_2})t$,
$U_{m_1m_2}=(1-\delta_{m_1,m_2})U$, and $V_{km}$ is $V_{k1}$ when $k\in L$ and
$V_{k2}$ when $k\in R$.

{\em Numerical results.}
We start with model (a) - single electronic level coupled to a molecular vibration (Fig.~\ref{fig1}a).
Electron escape rates to contacts are taken $\Gamma_L=\Gamma_R=0.1$~eV.
The frequency of the molecular vibration is set to $\omega_0=0.1$~eV and for the electron-vibration
coupling we take $U=0.05$~eV. The contacts temperature is taken as $T=300$~K. 
The Fermi energy is chosen as the energy origin $E_F=0$.
We apply a bias $V_{sd}=3$~V across the junction symmetrically ($\mu_L=1.5$~eV and
$\mu_R=-1.5$~eV) and consider the steady state of the system when level $\epsilon$ is moved in
and out of the bias window.
Calculations are performed on energy grid spanning the range from $-4$ to $4$~eV
with step $10^{-4}$~eV. 
We compare the results of zero order simulation, where rate (\ref{gamma_el}) and population (\ref{N_w0})
are obtained utilizing the zero order electron Green function in (\ref{Pi}), with SCBA results,
where the self-consistent procedure takes into account the mutual influence of electron and vibrational
degrees of freedom in the system is taken into account. In the latter case convergence is assumed
to be reached when difference in values of electron density matrix at subsequent steps is
less than $10^{-12}$.

\begin{figure}[htbp]
\centering\includegraphics[width=0.7\linewidth]{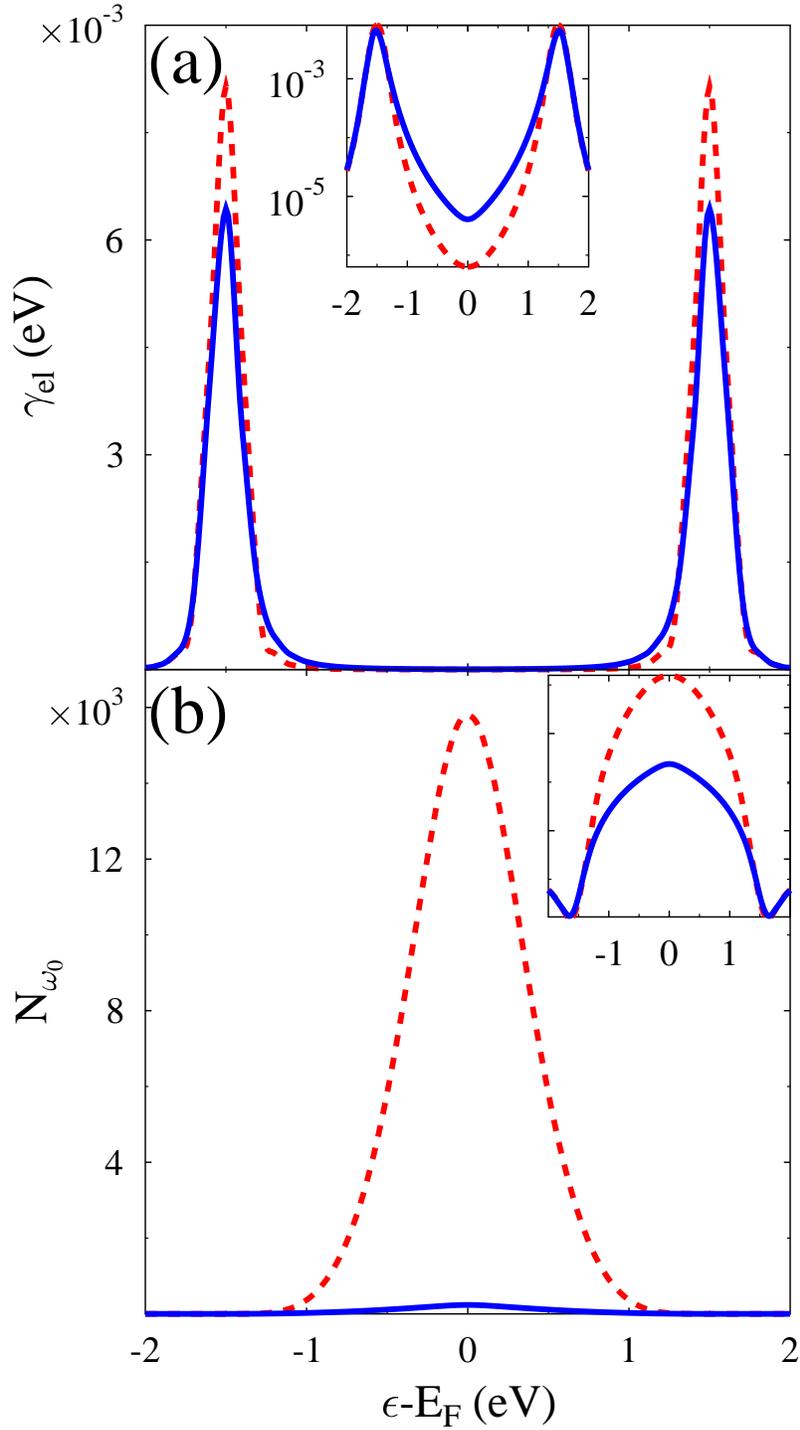}
\caption{\label{fig2}
Steady-state simulation of the single level model (Fig.~\ref{fig1}a)
within the SCBA (solid line, blue) and Born approximation (dashed line, red) schemes at $V_{sd}=3$~V.
Shown are (a) phonon dissipation rate due to coupling to electrons, Eq.~(\ref{gamma_el}),
and (b) nonequilibrium phonon population, Eq.~(\ref{N_w0}),
vs. position of the level $\epsilon$. 
Insets are logarithmic scale plots of the main panels.
}
\end{figure}

Figure~\ref{fig2}a shows phonon dissipation rates as function of gate voltage.
As expected, the rate is maximum when level $\epsilon$ crosses the lead chemical potential
where the possibility of effective creation of electron-hole pairs exceeds that of destruction
which leads to strong dissipation of vibrational energy; 
the rate is much lower away from chemical potentials  where both creation and destruction
of electron-hole pairs have similar probability.
Qualitatively both schemes give the same behavior. However, self-consistency of the SCBA
allows to account for multiple phonon scatterings, which results in significantly higher
dissipation rate for the vibration within the bias window.
As a result, the standard kinetic scheme significantly overestimates heating of molecular vibration
by electron flux, as is demonstrated in Fig.~\ref{fig2}b. This results in underestimation
of stability of molecular junction when analyzed within kinetic scheme.

Discrepancy between SCBA and standard kinetic scheme is even more pronounced for 
non-adiabatic electron-phonon coupling (model b).
This is the two-level model (Fig.~\ref{fig1}b)
used in Refs.~\cite{BrandbygePRL11,SegalPCCP12,FortiVazquezJPCL18} to demonstrate 
bias induced vibrational instabilities.
As above, we consider stable steady-state 
and its characteristics - rate (\ref{gamma_el}) and population (\ref{N_w0}).
We note that phonon back action on the electron degrees of freedom, characterized by 
the self-energy (\ref{Sigma}), is proportional to population $N_{\omega_0}$. 
That is, within the harmonic oscillator model, 
any electron pumping can be compensated by phonon back action when big enough
$N_{\omega_0}$ is reached. Thus, one expects that 
a stable steady state will be always achievable, and no phonon runaway will be observed.
Taking into account that molecular vibrations are not harmonic at high excitations,
the reasonable question to ask is if $N_{\omega_0}$ compensating for electronic pumping
is big enough to actually break molecular bond. 
We note that SCBA analysis of the model was performed in the literature previously~\cite{RyndykCunibertiPRB06,ParkMG_FCS_PRB11}.
Our goal here is comparison between the SCBA and kinetic scheme predictions.

The electronic levels are chosen at equilibrium as $\epsilon_1=-0.15$~eV and $\epsilon_2=0.55$~eV.
Following Ref.~\cite{BrandbygePRL11} we assume that the two electronic levels are pinned to 
their respective baths, so that positions of the levels are shifted with bias together with corresponding 
chemical potentials. 
Electron escape rates to contacts are $\Gamma_L=0.3$~eV and $\Gamma_R=0.1$~eV.
The other parameters are as in Fig.~\ref{fig2}.

\begin{figure}[htbp]
\centering\includegraphics[width=0.48\linewidth]{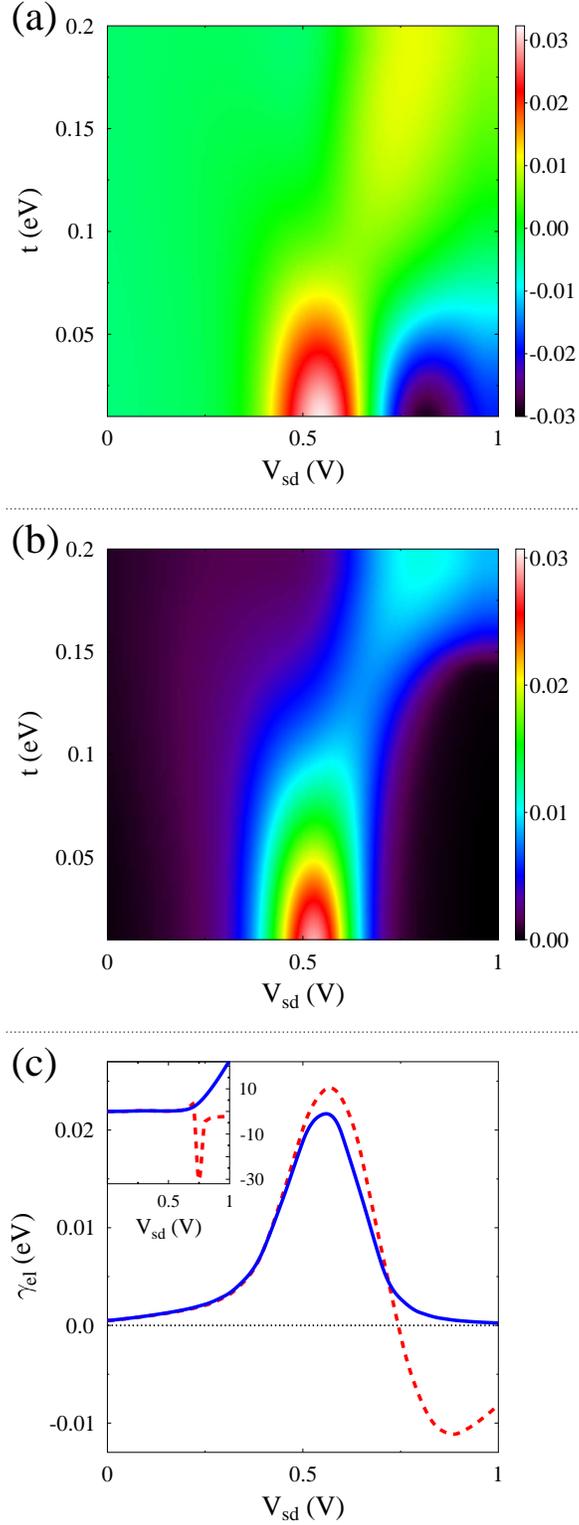}
\caption{\label{fig3}
Steady-state simulation of the two-level model (Fig.~\ref{fig1}b).
Shown is map of the phonon dissipation rate, Eq.~(\ref{gamma_el}),
vs. applied bias, $V_{sd}$, and electron hopping, $t$, for (a) zero order and (b) SCBA simulations.
Panel (c) shows results of the two simulations at $t=0.05$~eV.
Red dotted line presents kinetic scheme results, solid blue line - SCBA results.
Inset shows average vibrational population $N_{\omega_0}$, Eq.~(\ref{N_w0}), 
as function of $V_{sd}$.
}
\end{figure}

Figure~\ref{fig3} compares zero order and SCBA results for the phonon dissipation rate $\gamma_{el}$,
Eq.~(\ref{gamma_el}). In agreement with previous considerations~\cite{BrandbygePRL11,SegalPCCP12,FortiVazquezJPCL18},
the zero-order calculation predicts instability for resonance condition, showing runaway heating 
of the vibration when the electron hopping matrix element $t$ is small  (see low right corner of
the dissipation rate map shown in Fig.~\ref{fig3}a).
The corresponding SCBA results are shown in Fig.~\ref{fig3}b: no instability
(negative dissipation rate) is observed in this case. 
To make the comparison easier, Fig.~\ref{fig2}c shows horizontal cuts of the two maps 
for $t=0.05$~eV. The inset in this panel shows the nonequilibrium population of the mode at this time. 
While the zero-order simulation predicts negative damping (and hence instability), 
SCBA result indicates finite heating of the mode with bias. 
Note that the population at $V_{sd}=1$~V is about $N_{\omega_0}=20$, 
which for $\omega_0=0.1$~eV gives total vibrational energy of $2$~eV ($190$~kJ/mol) --
insufficient for breaking most molecular bonds.

The qualitative nature of this results, that is, the absence of true instability in the models considered, 
does not depend on the parameters used in the calculation. 
We note that phonon back action on the electron degrees of freedom, characterized by the self-energy (5), 
is proportional to population $N_{\omega_0}$.
That is, within the harmonic oscillator model, any electron pumping can be compensated by phonon 
back action when big enough $N_{\omega_0}$ is reached. Thus, one expects that a stable steady 
state will be always achievable, and no phonon runaway will be observed.
Depending on the actual molecular forcefield, the corresponding   compensating for electronic pumping 
may be quite large~\cite{schinabeck_hierarchical_2018} and, depending on the molecule, 
may be beyond the bond-breaking threshold of the real anharmonic molecule. 
Such bond-breaking should not however be deduced just from the prediction of negative 
vibrational dissipation rate obtained from the standard kinetic analysis.
Note that negativity of vibrational dissipation rate
in a steady-state situation is an indication of qualitative failure of 
the zero order treatment.

\begin{figure}[htbp]
\centering\includegraphics[width=0.7\linewidth]{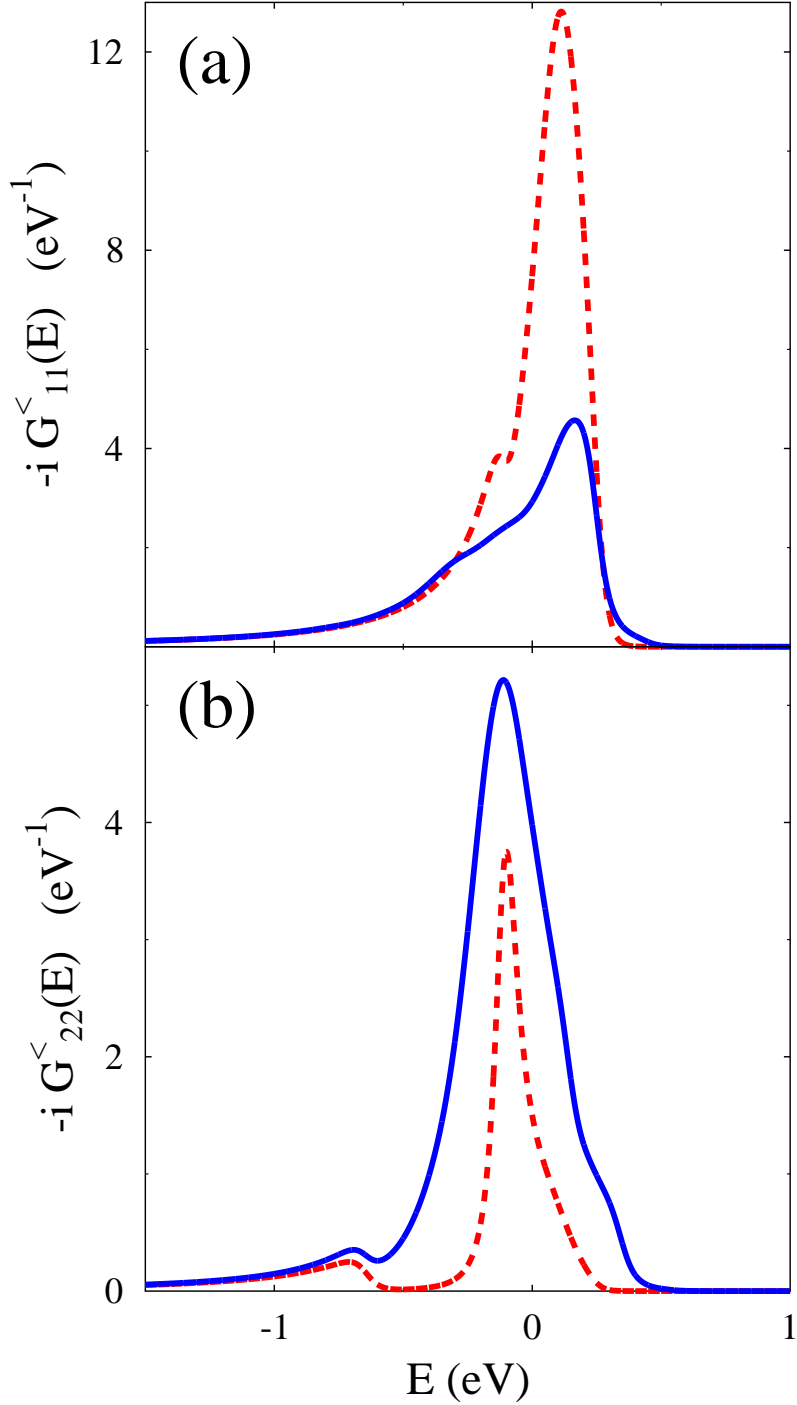}
\caption{\label{fig4}
Steady-state simulation of the two-level model (Fig.~\ref{fig1}b).
Shown are electron population distributions for levels (a) $\epsilon_1$ and 
(b) $\epsilon_2$. Kinetic scheme results (dashed line, red) are compared
with the SCBA (solid line, blue) simulations. 
}
\end{figure}

Figure~\ref{fig4} shows the electronic energy distribution in levels (a) $\epsilon_1$ 
and (b) $\epsilon_2$ calculated with (solid line - SCBA) and without (dashed line - zero order)
vibrational back action taken into account. 
In this calculation we have used $t=0.05$~eV and $V_{sd}=0.9$~V.
These parameters correspond to the most unstable (most negative dissipation rate)
prediction of the Born (zero order) calculation (Fig.~\ref{fig3}c).
One sees that electron-vibration coupling promotes redistribution of electron population 
between levels $\epsilon_1$ and $\epsilon_2$; the effect is significant even for $U\ll\Gamma_{L,R}$.
We note in passing that effect of the coupling on electronic coherence (not shown) 
is even more drastic. 

{\em Conclusions.}
Standard rate theories that are very useful in the analysis of many chemical dynamics phenomena, 
usually rely on timescale separation between the system of interest and its environment. 
Failure of such separation in treatments of systems interacting with equilibrium environments is usually 
handled by redefining the boundaries between system and bath. Extra care is needed when the system 
is driven by a non-equilibrium environment, where the driving may move the system into regimes 
where timescale separation does not hold. 
We have discussed the implications of the common timescale separation assumption used in 
analyzing the time evolution vibrational energy in biased molecular junctions. 
Using such treatments outside their range of validity can lead to qualitatively wrong predictions.
As an example, we have consider generic models of molecular junctions with electron-phonon interaction
treated within the NEGF-SCBA level of theory. 
Standard timescale separation argument suggests that phonon back-action on electronic degrees of 
freedom can be disregarded. Such approximation, however, formally violates conservation  laws 
and can fail both qualitatively and quantitatively when inadvertently carried into regimes where timescale separation does not hold.
Not accounting for this back action leads to an overestimated heating of molecular vibrations in 
the standard single electronic level model of current carrying molecular junctions as compared with 
the renormalized (SCBA) treatment (Fig.~\ref{fig1}a).
This discrepancy  with the SCBA is even more pronounced for non-adiabatic electron-phonon coupling 
model (Fig.~\ref{fig1}b). 
Analysis of this model within the timescale separation assumption  has indicated 
the existence of bias induced vibrational instability in molecular junctions,
which was associated with appearance of negative vibrational dissipation rate.
However, a self-consistent calculation, here carried at the NEGF-SCBA level, shows
that stable steady-state is reached for any set of parameters (any electronic heating rate).

Depending on the molecular forcefield, the molecule-metal coupling and potential bias,
the molecular energy at the steady state obtained in such a (harmonic model) calculation, 
which can be high~\cite{schinabeck_hierarchical_2018}, may or may not exceed the actual 
bond-breaking threshold of the real anahrmonic molecule.
We note that sudden changes in electronic system (such as fast switch on of bias) 
can lead to transient heating spikes that, for a harmonic oscillator, will eventually relax to 
the new steady state but in real molecules can lead to bond breaking even if 
the steady-state population is below the breaking threshold.
For slow switch-on of the bias, observation of vibrational instabilities in calculation done under 
the standard system-bath timescale separation assumption should be taken as indications 
that this assumption fails and that higher-level studies are needed for reaching conclusions 
about actual bond-breaking.

Rate theories using standard kinetic schemes are often a method of choice that has been repeatedly 
reliable and useful for modeling chemical dynamics. Extra caution should be exercised when 
employing such methods in nonequilibrium systems, since they usually disregard back action of 
the system onto its bath(s) which, as we showed, may lead to erroneous predictions.
Development of advanced kinetic schemes for the latter systems is a goal of future research.

\begin{acknowledgement}
A.N. is supported by the National Science Foundation (Grant No. CHE-1665291), 
the Israel Science Foundation, the US-Israel Binational Science Foundation,
and the University of Pennsylvania.
M.G. is supported by the National Science Foundation under CHE-1565939
and by the Department of Energy under DE-SC0018201.
\end{acknowledgement}

\providecommand{\latin}[1]{#1}
\makeatletter
\providecommand{\doi}
  {\begingroup\let\do\@makeother\dospecials
  \catcode`\{=1 \catcode`\}=2 \doi@aux}
\providecommand{\doi@aux}[1]{\endgroup\texttt{#1}}
\makeatother
\providecommand*\mcitethebibliography{\thebibliography}
\csname @ifundefined\endcsname{endmcitethebibliography}
  {\let\endmcitethebibliography\endthebibliography}{}

\end{document}